\documentclass[a4paper,fleqn,usenatbib,letters]{mnras}

\usepackage{newtxtext,newtxmath}
\usepackage[T1]{fontenc}
\usepackage{ae,aecompl}

\usepackage{graphicx}	% Including figure files
\usepackage{amsmath}	% Advanced maths commands
\usepackage{amssymb}	% Extra maths symbols

\newcommand\degrees[1]{\ensuremath{#1^\circ}}
\newcommand{\as}{$^{\prime\prime}$}

\title[]{Dipper disks not inclined towards edge-on orbits}

\author[Ansdell et al.]{M. Ansdell,$^{1}$ E. Gaidos,$^{2,3}$ J. P. Williams,$^{1}$ G. Kennedy,$^{4}$ M. C. Wyatt,$^{4}$
\newauthor D. M. LaCourse,$^{5}$ T. L. Jacobs,$^{5}$ A. W. Mann$^{6}$ \\
$^1$Institute for Astronomy, University of Hawai`i at M\={a}noa, Honolulu, HI, USA \\
$^2$Department of Geology \& Geophysics, University of Hawai`i at M\={a}noa, Honolulu, HI, USA \\
$^3$Center for Space and Habitability, University of Bern, Bern, Switzerland \\
$^4$Institute of Astronomy, University of Cambridge, Madingley Road, Cambridge, UK \\
$^5$Amateur Astronomer \\
$^6$Hubble Fellow, Department of Astronomy, The University of Texas at Austin, Austin, TX, USA
}

\date{Accepted July 12, 2016}

\pubyear{2016}

\begin{document}
\label{firstpage}
\pagerange{\pageref{firstpage}--\pageref{lastpage}}
\maketitle

% Abstract of the paper
\begin{abstract}
The so-called ``dipper" stars host circumstellar disks and have optical and infrared light curves that exhibit quasi-periodic or aperiodic dimming events consistent with extinction by transiting dusty structures orbiting in the inner disk. Most of the proposed mechanisms explaining the dips---i.e., occulting disk warps, vortices, and forming planetesimals---assume nearly edge-on viewing geometries. However, our analysis of the three known dippers with publicly available resolved sub-mm data reveals disks with a range of inclinations, most notably the face-on transition disk J1604-2130 (EPIC~204638512). This suggests that nearly edge-on viewing geometries are {\em not} a defining characteristic of the dippers and that additional models should be explored. If confirmed by further observations of more dippers, this would point to inner disk processes that regularly produce dusty structures far above the outer disk midplane in regions relevant to planet formation.
\end{abstract}

% Select between one and six entries from the list of approved keywords.
% Don't make up new ones.

%%%%%%%%%%%%%%%%%%%%%%%%%%%%%%%%%%%%%%%%%%%%%%%%%%
%%%%%%%%%%%%%%%%% BODY OF PAPER %%%%%%%%%%%%%%%%%%
%%%%%%%%%%%%%%%%%%%%%%%%%%%%%%%%%%%%%%%%%%%%%%%%%%

%======================= INTRODUCTION =========================

\begin{table*}
\begin{center}
\caption{Properties of Dippers with Resolved Sub-mm Images}
\begin{tabular}{lllllccccccc}
EPIC & 2MASS & RA$_{\rm J2000}$ & Dec$_{\rm J2000}$ & Mem. & SpT & $T_{\rm eff}$ (K) & $A_{\rm V}$ (mag) & Disk & $P_{\rm rot}$ (d) & $D_{\rm dip}$ & Var.   \\
\hline
204638512  &  16042165-2130284  &  16:04:21.655  &  -21:30:28.50  &  USc  &  K2   &  4900  &  0.7  &  TD   & 5.00  &  0.57  &  A   \\              
205151387  &  16090075-1908526  &  16:09:00.762  &  -19:08:52.70  &  USc  &  M1  &  4000  &  0.8  &  Full  &  9.55  &  0.31  &  Q  \\
203850058  &  16270659-2441488  &  16:27:06.596  &  -24:41:48.84  &  Oph  &  M6  &  2700  &  3.0  &  Full  &  2.88  &  0.18  &  Q   \\
\hline
\end{tabular}
\label{tab-dippers}
\end{center}
\end{table*}

\begin{figure*}
\begin{centering}
\includegraphics[width=17cm]{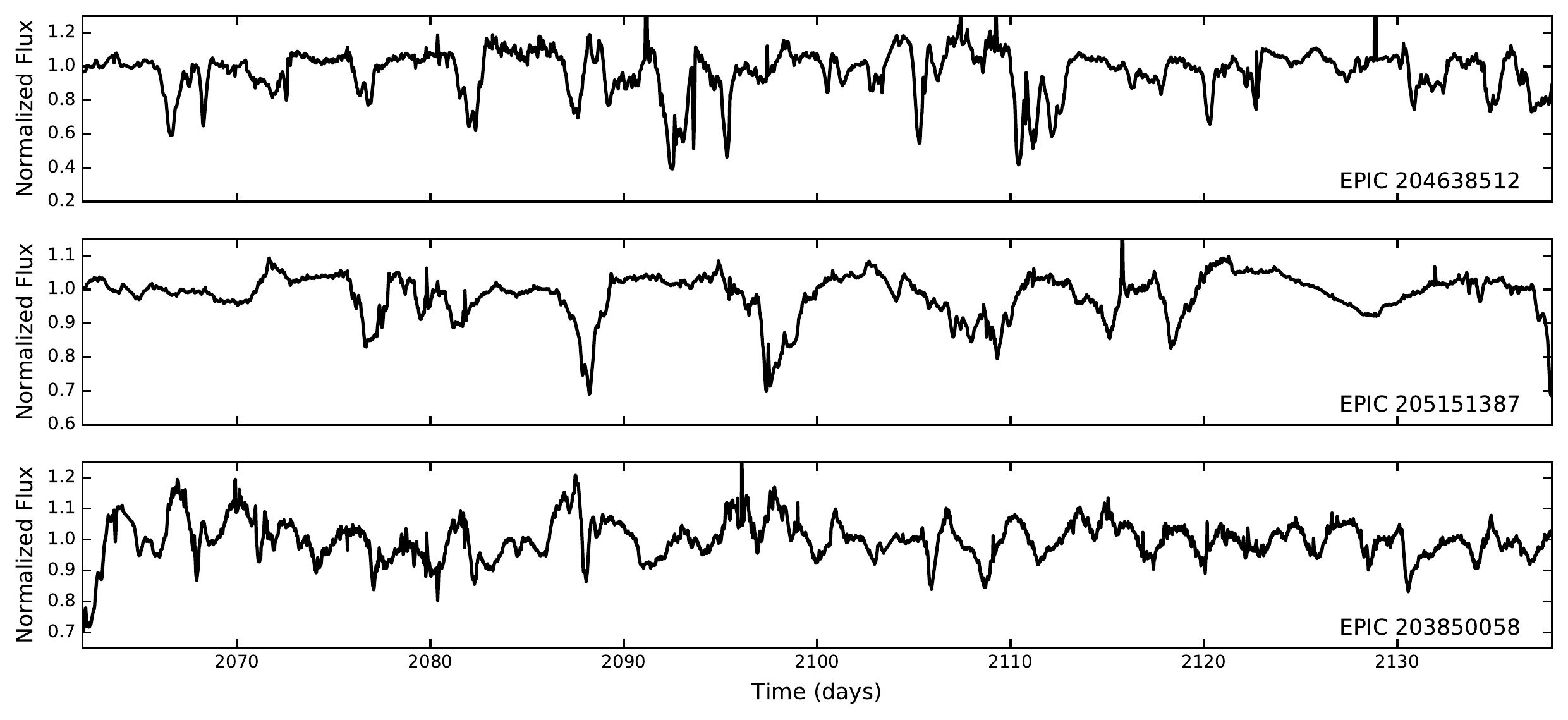}
\caption{\small Normalized K2 light curves (Section~\ref{sec-k2}) showing $\gtrsim$10\% dip depths with $\sim$0.5--2 day durations typical of dipper stars.}
\label{fig-lcs}
\end{centering}
\end{figure*}

\section{Introduction}

Understanding how planets form is one of the most compelling problems in astronomy. Close-in planets appear to be common \citep{2010Sci...330..653H,2013ApJ...770...69P,2015ApJ...799..180S} and we can study their nascent systems (i.e., protoplanetary disks) in young stellar associations. However, observing planet formation at $\lesssim1$~AU is complicated by small angular scales and faint disk emission compared to the host star; for the nearest star-forming regions, the best achievable angular resolution for optical/infrared scattered light and sub-mm images is only a few AU. 

However, new probes of the inner disk may be the so-called ``dipper" stars, whose optical and infrared light curves exhibit episodic drops in flux consistent with extinction by transiting dusty structures orbiting with Keplarian periods of down to a few days. Dipper stars were identified with the CoRoT and {\it Spitzer} missions in the young ($\sim$2--3~Myr) Orion Nebula Cluster \citep{2011ApJ...733...50M} and NGC~2264 region \citep{2010A&A...519A..88A,2014AJ....147...82C}. These studies found that the depth, duration, and periodicity of the dips are consistent with extinction by dust orbiting near the star-disk co-rotation radius. \cite{2015A&A...577A..11M} proposed that the dippers in NGC~2264 could be explained by occulting inner disk warps driven by accretion streams onto the host star, as reported for AA Tau \citep{1999A&A...349..619B}. Unfortunately, the significant distances to these clusters ($\sim$400-750~pc) limited follow-up observations.

\cite{2016ApJ...816...69A} identified $\sim$25 dippers in the young ($\lesssim10$~Myr), nearby ($\sim$120--145~Myr) Upper Sco and $\rho$ Oph star-forming regions using high-precision optical photometry from the K2 mission \citep{2014PASP..126..398H}. Follow-up observations showed that the K2 dippers are often weakly accreting late-type stars hosting moderately evolved primordial disks, challenging the disk warp scenario described above. Thus \cite{2016ApJ...816...69A} proposed alternative mechanisms to explain the dips, namely occulting vortices at the inner disk edge produced by the Rossby Wave Instability and transiting clumps of circumstellar material related to planetesimal formation. \cite{2016arXiv160503985B} also found that magnetospheric truncation of weakly accreting disks with misaligned magnetic fields could form occulting accretion streams that produce dippers with high to moderate inclinations.

The proposed mechanisms for explaining the dips therefore assume geometries that are closer to edge-on than face-on. The occurrence rate of dippers in co-eval clusters (e.g., $\sim$20--30\% of classical T Tauri stars in NGC~2264; \citealt{2010A&A...519A..88A}, \citealt{2014AJ....147...82C}) and the moderate optical extinction towards these objects \cite[$A_{\rm V}\lesssim1$;][]{2016ApJ...816...69A} suggest that we are not seeing the disks directly edge-on ($i=90$\degrees{}), but rather viewing the dippers at nearly edge-on inclinations (e.g., $i=70$\degrees{}) and thus observing transits of dusty structures lifted above the disk midplane. However, these geometric assumptions have not yet been compared to observations.

In this Letter, we present the three known dippers whose circumstellar disks have been resolved in archival sub-mm data such that their inclinations can be constrained. Surprisingly, we find disks with a range of inclinations, most notably the face-on transition disk (TD) J1604-2130 \citep{2012ApJ...753...59M,2014ApJ...791...42Z}. This indicates that nearly edge-on disk inclinations are {\em not} a defining characteristic of the dippers, and motivates a re-examination of the dipper mechanisms so that we can properly interpret these objects in the context of planet formation.

\begin{figure*}
\begin{centering}
\includegraphics[width=17cm]{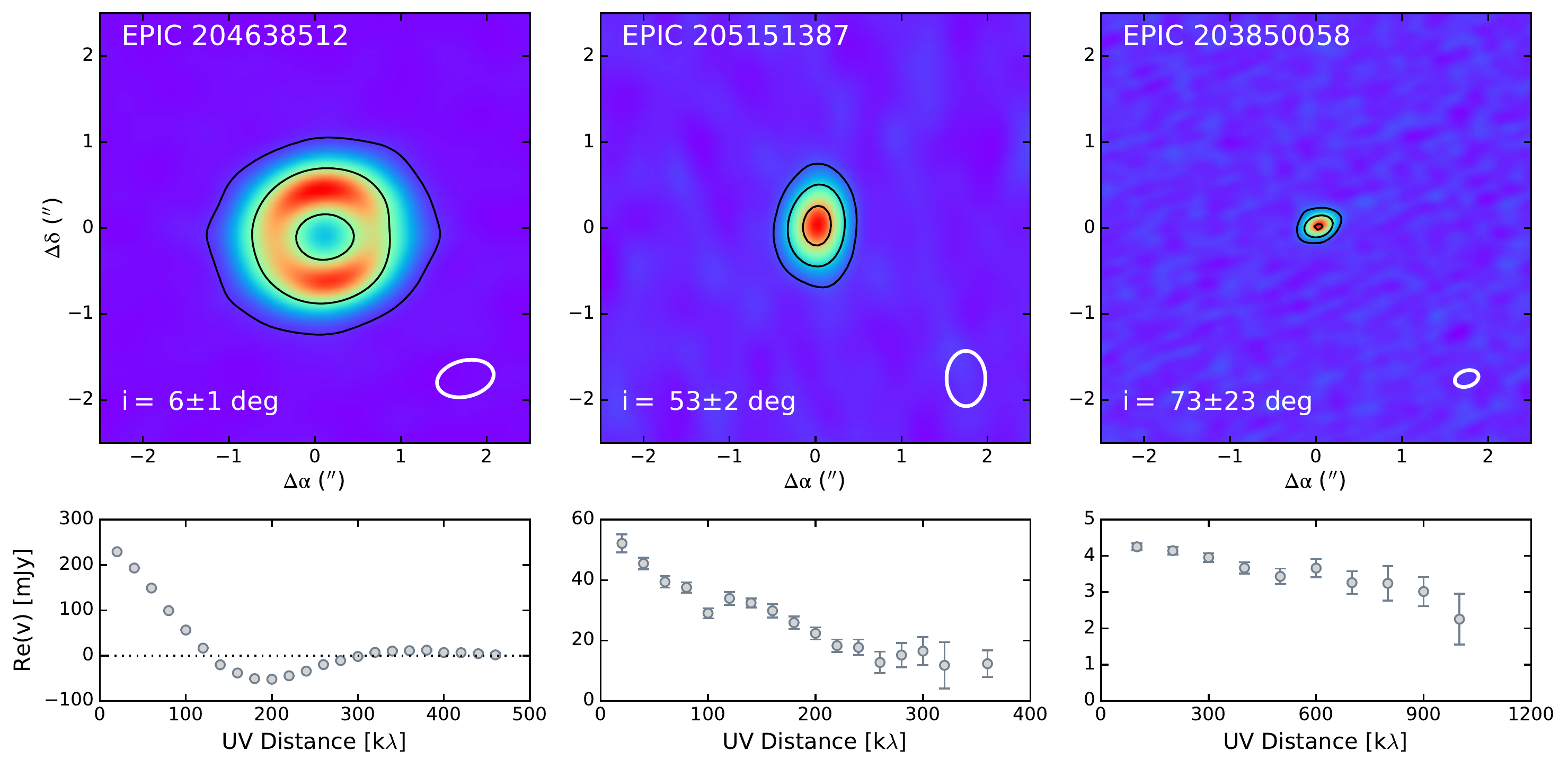}
\caption{\small Top panels show ALMA sub-mm continuum images (5\as{}$\times$5\as{}) with fitted disk inclinations and beam sizes (Section~\ref{sec-alma} \& \ref{sec-inclinations}). Contours are 10$\sigma$ and 100$\sigma$ for EPIC~204638512 and 5$\sigma$, 20$\sigma$, and 50$\sigma$ for EPIC~205151387 and EPIC~203850058. Bottom panels show the real part of the visibilities as a function of projected baseline length.}
\label{fig-alma}
\end{centering}
\end{figure*}

%=======================  SAMPLE =========================

\section{Sample}
\label{sec-sample}

Our sample consists of the three dippers with publicly available high-resolution sub-mm data sufficient to constrain disk inclinations. The dippers were identified from their K2 light curves and are located in the $\sim$10-Myr old Upper Sco \citep{2012ApJ...746..154P} and $\sim$1-Myr old $\rho$ Oph \citep{2007ApJ...671.1800A} star-forming regions. EPIC~205151387 and EPIC~203850058 were reported in \cite{2016ApJ...816...69A}, while EPIC~204638512 is a newly identified dipper. To our knowledge, these are the only currently known dippers with resolved sub-mm data. Table~\ref{tab-dippers} gives their Ecliptic Plane Input Catalog (EPIC) and Two Micron All-Sky Survey \cite[2MASS; ][]{2006AJ....131.1163S} names, right ascension and declination, cluster membership (Mem.), spectral type (SpT), effective temperature ($T_{\rm eff}$), optical extinction ($A_{\rm V}$), and disk type (Disk). Stellar properties are from the literature \citep{2002A&A...393..597N,2014ApJ...787...42C,2016ApJ...816...69A}.

EPIC~204638512 is a K-type pre-main sequence star with spectroscopic signatures of weak accretion \cite[e.g.,][]{2012ApJ...745...56D} and hosts the face-on TD known as J1604-2130 \citep{2012ApJ...753...59M,2014ApJ...791...42Z}. EPIC~205151387 is a pre-main sequence M1 star that hosts a full disk \citep{2014ApJ...787...42C} and exhibits variable accretion signatures \citep{2012ApJ...745...56D,2016ApJ...816...69A}. EPIC~203850058 is the brown dwarf $\rho$~Oph~102, which hosts a well-studied full disk \citep{2002A&A...393..597N,2012ApJ...761L..20R} that shows evidence for significant mass accretion as well as winds and molecular outflows \citep{2004A&A...424..603N,2005Natur.435..652W,2008ApJ...689L.141P,2010ApJS..188...75M,2015A&A...579A..66M}.

%======================= DATA ANALYSIS =========================

\section{DATA \& ANALYSIS}
\label{sec-data}
 
\subsection{K2 Optical Light Curves \label{sec-k2}}

Figure~\ref{fig-lcs} presents the K2 light curves of the three dippers in our sample. As in \cite{2016ApJ...816...69A}, we use the \textsc{K2SFF} light curves made publicly available by the Mikulski Archive for Space Telescopes (MAST). These light curves were extracted from the {\it Kepler} Target Pixel Files (TPFs) using photometric apertures with the Self Field Flattening (SFF) technique, which corrects for spacecraft motion by correlating observed flux variability with spacecraft pointing \citep{2014PASP..126..948V}.

For EPIC~204638512 and EPIC~205151387, fainter secondary sources are visible within their \textsc{K2SFF} photometric apertures. However, these potentially contaminating interlopers are at sufficiently large projected distances (\textgreater8\as) from the primary targets that their separate light curves can be manually extracted. We obtained the TPFs from MAST and used the {\sc KepPCA} pipeline to apply custom photometric apertures for individually extracting the light curves of the primary and secondary sources while also correcting for spacecraft motion. We confirmed that the dipper activity is associated with the primary target with negligible contributions from the secondary source in both cases. We use the \textsc{K2SFF} light curves in the remainder of this work, as they show improved pointing error corrections compared to the light curves produced by our \textsc{KepPCA} pipeline.

We used the K2 light curves to infer stellar rotation periods ($P_{\rm rot}$) and characterize the dipper activity in terms of dip depth ($D_{\rm dip}$) and quasi-periodic (Q) or aperiodic (A) variability (see Section 2.2 in \citealt{2016ApJ...816...69A}). Table~\ref{tab-dippers} provides these dipper properties, though the $P_{\rm rot}$ of EPIC~204638512 is uncertain due to a weak rotational signal. The three dippers in our sample generally follow the trends with stellar and disk properties identified in \cite{2016ApJ...816...69A}, indicating that they are not unusual among the overall dipper population. In particular, they follow the correlation between extinction-corrected WISE-2 excess and $D_{\rm dip}$ \cite[see Figure 9 in][]{2016ApJ...816...69A}, which was interpreted as evidence that the dips are related to hot inner (rather than cool outer) disk material. The three dippers also follow the observed correlation between $P_{\rm rot}$ and $T_{\rm eff}$, although EPIC~205151387 has a long $P_{\rm rot}$ for its $T_{\rm eff}$ value. SED modeling by \cite{2016ApJ...816...69A} also showed that dippers typically have inner dust disks extending to within a few stellar radii. The infrared color excesses for EPIC~205151387 and EPIC~203850058 indicate they host full disks \cite[see Figure 6 in][]{2016ApJ...816...69A}. Although EPIC~204638512 hosts a disk with a large inner cavity in the sub-mm, it exhibits variable infrared excess consistent with dust at small ($\lesssim$0.1~AU) orbital radii \citep{2010AJ....140.1444D,2014ApJ...791...42Z}.

\subsection{Resolved Sub-mm Data \label{sec-alma}}

EPIC~204638512 hosts the well-characterized face-on TD known as J1604-2130, which was discovered by \cite{2012ApJ...753...59M} using the Sub-millimeter Array (SMA; \citealt{2004ApJ...616L...1H}). \cite{2012ApJ...753...59M} resolved a $\sim$70-AU inner dust cavity in their continuum map and derived a precise disk inclination of 6\degrees{}$\pm1.5$\degrees{} using their $^{12}$CO 3--2 first-moment map. \cite{2014ApJ...791...42Z} later compared the dust and gas radial structures using ALMA Cycle~0 observations; Figure~\ref{fig-alma} shows the cleaned 880~$\mu$m continuum image from the ALMA Science Archive (project code: 2011.0.00526.S) with a clearly resolved source and 0.67\as$\times$0.42\as ($\sim$100$\times$60~AU) beam.

EPIC~205151387 was also observed during ALMA Cycle~0 (project code: 2011.0.00526.S) in the 880~$\mu$m continuum and $^{12}$CO 3--2 line \citep{2014ApJ...787...42C}. Figure~\ref{fig-alma} shows the cleaned continuum image from the ALMA Science Archive with a 0.64\as$\times$0.45\as ($\sim$95$\times$65~AU) beam. We also plot the real part of the visibilities as a function of projected baseline length, where the clear decrease in visibility with UV distance indicates a spatially resolved source.

EPIC~203850058 was imaged at high resolution in the 870~$\mu$m continuum during ALMA Cycle~1 (project code: 2012.1.00046.S). The calibrated continuum data were not available from the ALMA Science Archive, thus we downloaded the raw visibilities and executed the pipeline calibration script, then extracted the continuum image by averaging over the continuum channels and interactively cleaning with a Briggs robust weighting parameter of $+0.5$, giving a beam size of 0.28\as$\times$0.19\as~($\sim$35$\times$25~AU). Figure~\ref{fig-alma} shows the cleaned continuum image and the real part of the visibilities as a function of projected baseline length; the slight decrease in the visibilities indicates a marginally resolved source.

\subsection{Adaptive Optics Imaging\label{sec-ao}}

We searched the literature for adaptive optics (AO) imagery to check whether close companions could be influencing the dips. EPIC~204638512 has been imaged extensively with AO, providing strict limits on close companions. \cite{2008ApJ...679..762K} used aperture masking interferometry combined with direct imaging to rule out the presence of close companions from $\sim$0.07~$M_{\odot}$ at $\sim$2~AU to $\sim$0.02~$M_{\odot}$ at $\sim$40~AU. \cite{2011ApJ...726..113I} used direct imaging to limit companions from $\sim$0.07~$M_{\odot}$ at $\sim$60~AU to $\sim$0.005~$M_{\odot}$ at $\gtrsim$300~AU. High-resolution optical spectroscopy showed no evidence for a double-line spectroscopic binary \citep{2012ApJ...745...56D}.

EPIC~205151387 also has strict limits on close companions from extensive AO imaging. Companions are ruled out from $\sim$0.03~$M_{\odot}$ at $\sim$2~AU to $\sim$0.01~$M_{\odot}$ at $\sim$40~AU \citep{2008ApJ...679..762K} and from $\sim$0.10~$M_{\odot}$ at $\sim$60~AU to $\sim$0.008~$M_{\odot}$ at $\gtrsim$300~AU \citep{2011ApJ...726..113I}. High-resolution optical spectroscopy showed no signs of a double-line spectroscopic binary \citep{2012ApJ...745...56D}. 

EPIC~203850058 is an ultra-cool brown dwarf, making it a particularly difficult target for AO imaging. We do not know of any existing AO data for this source.

%======================= DISCUSSION =========================

\section{Discussion}
\label{sec-discussion}
 
\subsection{Disk Inclinations\label{sec-inclinations}}

For EPIC~204638512, we adopt the disk inclination of $i=6$\degrees{}$\pm$1.5\degrees{} derived by \cite{2012ApJ...753...59M}, who placed strong constraints on disk geometry using their $^{12}$CO first-moment map and assumptions of Keplerian rotation. For EPIC~205151387 and EPIC~203850058, we derive disk inclinations from their sub-mm continuum data (Section~\ref{sec-alma}) using standard routines from the {\it Common Astronomy Software Applications} (CASA) package \citep{2007ASPC..376..127M}; although their existing CO data were insufficient to derive precise disk inclinations, the first-moment maps can be used as rough checks on our continuum results.

We derived the disk inclination of EPIC~205151387 using the CASA routine \texttt{uvmodelfit}, which fits simple analytic source component models (point-source, Gaussian, or disk) directly to the visibility data. We assumed an elliptical Gaussian model, which has six free parameters: integrated flux density ($F$), FWHM along the major axis ($a$), aspect ratio of the axes ($r$), position angle (PA), right ascension offset from the phase center ($\Delta\alpha$), and declination offset from the phase center ($\Delta\delta$). We found $F=49.5\pm0.4$~mJy and PA$=-22$\degrees{}$\pm$3\degrees{}, then derived the inclination from $r$ assuming circular disk structure, finding $i=53$\degrees{}$\pm$2\degrees{}. To check our results, we analyzed the first-moment $^{12}$CO map, finding a similar position angle (${\rm PA}\approx-30$\degrees{}) and inclination ($i\approx50$\degrees{}) when assuming Keplerian rotation.

EPIC~203850058 is thought to have a high disk inclination based on the detection of an optical jet \citep{2005Natur.435..652W} and the nearly symmetric morphology of a bipolar outflow in molecular CO \citep{2008ApJ...689L.141P}, as previously noted by \cite{2012ApJ...761L..20R}. To derive an inclination, we again used \texttt{uvmodelfit} to fit an elliptical Gaussian model to the continuum visibility data, but with an initial guess of ${\rm PA}\approx20$\degrees{} based on the first-moment $^{12}$CO map in \cite{2012ApJ...761L..20R}. We found $F=4.0\pm0.1$~mJy and PA$=15$\degrees{}$\pm$1\degrees{}, consistent with \citealt{2012ApJ...761L..20R}. We then derived the inclination from $r$ assuming circular disk structure, finding $i\approx90$\degrees{}, but with very large errors. We therefore checked our results with the CASA routine \texttt{imfit}, which fits an elliptical Gaussian to the source in its image plane, then uses the clean beam to return de-convolved fit results; we found a nearly edge-on disk with $i=73$\degrees{}$\pm$23\degrees{}, $F=4.2\pm0.1$~mJy, and ${\rm PA}=$10\degrees{}$\pm$15\degrees{}. Although the small size and faint emission of this source complicates the analysis, the overall picture appears to point to a nearly edge-on viewing geometry for EPIC~203850058.

%To check the reliability of these results, we created a model elliptical Gaussian disk with the same size, flux, and inclination of EPIC~203850058, then used \texttt{simalma} in CASA to simulate ALMA observations with the same sensitivity and resolution as the real observations; we recovered the inclination with comparable errors ($i=74$\degrees{}$\pm$16\degrees{}). Although our inclination uncertainties are large (likely due to the disk being marginally resolved), \cite{2012ApJ...761L..20R} noted additional evidence for a high inclination, namely the detection of an optical jet \citep{2005Natur.435..652W} and the nearly symmetric morphology of a bipolar outflow in molecular CO \citep{2008ApJ...689L.141P}.

Note that these sub-mm observations have resolutions of $\sim$20--50~AU in radius (Section~\ref{sec-alma}), thus the estimated inclinations reflect bulk disk geometry and assume a uniform inclination angle throughout the disk. Moreover, the uncertainties do not include systematic errors (e.g., we assume the observed disks of EPIC 205151387 and EPIC 203850058 are adequately represented by elliptical Gaussians).

\subsection{A Call For Re-thinking Dipper Mechanisms\label{sec-mechanisms}}

The proposed mechanisms explaining the dipper phenomenon favor geometries that are nearly edge-on. The disks are likely not seen completely edge-on ($i=90$\degrees{}), however, due to the moderate extinction towards the dippers \cite[$A_{\rm V}\lesssim1$;][]{2016ApJ...816...69A}. Rather, it is thought that we are viewing the dippers at nearly edge-on inclinations (e.g., $i=70$\degrees{}) and thus observing transits of occulting material lifted above the disk midplane by some process \cite[e.g., the breakdown of Rossby waves into vortices;][]{2016ApJ...816...69A}. Notably, none of the proposed dipper mechanisms can account for obscurations from face-on disks. 
 
Thus the surprising range of dipper disk inclinations presented in this work, in particular the face-on geometry of EPIC~204638512 (J1604-2130), suggests that nearly edge-on viewing geometries are {\em not} a defining characteristic of the dippers and motivates the exploration of alternative models (or combinations of models) that can explain a range of disk inclinations. For example, occulting accretion streams \cite[e.g.,][]{2015A&A...577A..11M,2016arXiv160503985B} could possibly account for even face-on outer disks if they act in concert with other mechanisms warping the inner disk, such as  dynamical interactions with (proto-) planets or low-mass stellar companions \cite[e.g.,][]{2014MNRAS.442.3700F,2015ApJ...798L..44M}. Populations of scattered planetesimals from migrating (proto-) planets may also explain low dipper disk inclinations \citep{2011A&A...531A..80K}, but more work is needed to explore nearly polar orbits.

%As a simple test, we calculated the log likelihood ratio where one model assumes an isotropic inclination distribution and the other model constrains disk inclination to greater than some cutoff value. We used the observed inclinations presented in this work and ran Monte Carlo simulations to sample the measurement errors assuming normal distributions. We found that the likelihood ratio approaches unity as the cutoff inclination approaches zero, and that any restriction in inclination is \textless20\degrees{}, consistent with an isotropic distribution of disk inclinations.

\subsection{EPIC~204638512 (J1604-2130) \label{sec-epic512}}

EPIC~204638512 is a particularly interesting case. This source hosts a face-on disk ($i=6$\degrees{}$\pm$1.5\degrees{}; \citealt{2012ApJ...753...59M}) with a large sub-mm dust cavity ($\sim$80~AU in radius; \citealt{2014ApJ...791...42Z}), which seemingly makes it an unlikely dipper. Yet, EPIC~204638512 exhibits the deepest flux dips among the known K2 dippers (up to $\sim$60\%; Figure~\ref{fig-lcs}). How can these characteristics be reconciled?

The dipper activity may be related to an inclined and variable inner dust disk, as implied from its infrared emission. The object's {\it Spitzer} IRAC photometry shows no excess \citep{2012ApJ...753...59M}, while its {\it Spitzer} IRS spectrum and WISE photometry reveal excesses consistent with dust at small ($\lesssim$0.1~AU) orbital radii \citep{2010AJ....140.1444D,2014ApJ...791...42Z}. A factor of four variability in mid-infrared flux was also seen over several weeks, indicating a rapidly changing inner dust disk \citep{2009AJ....137.4024D}. Moreover, \cite{2014ApJ...795...71T} used near-infrared imaging polarimetry to identify intensity nulls in the outer disk annulus, which could be self-shadowing from a misaligned inner disk.

An inclined transient inner disk has been proposed for HD~142527, which also hosts a face-on transition disk with a large inner dust gap \citep{2006ApJ...636L.153F} and exhibits intensity nulls along the outer disk annulus in its infrared scattered light images \citep{2012ApJ...754L..31C}. HD~142527 has a known inner disk, thought to be a transient feature of accretion from the outer disk \citep{2011A&A...528A..91V,2013Natur.493..191C}. \cite{2015ApJ...798L..44M} modeled the system, finding a relative inclination of $\sim$70\degrees{} between the inner and outer disks, possibly due to dynamical interactions with a low-mass stellar companion orbiting inside the dust gap \citep{2012ApJ...753L..38B,2014ApJ...791L..37R,2014ApJ...781L..30C}. 

A similar scenario for EPIC~204638512 would reconcile its dips and face-on outer disk. One indication of an inclined inner disk is EPIC~204638512's weak rotational signal (Section~\ref{sec-k2}), which suggests the star is pole-on and thus aligned with the outer disk. The dust cavity of EPIC~204638512 is also thought to have been cleared by giant planet(s) orbiting inside the dust gap \citep{2012ApJ...753...59M,2014ApJ...791...42Z,2015A&A...579A.106V}, which could drive an inner disk warp. However, giant planets alone likely cannot account for all the dippers, as dipper occurrence rates (e.g., $\sim$20--30\% in NGC~2264; \citealt{2010A&A...519A..88A}, \citealt{2014AJ....147...82C}) are much larger than giant planet occurrence rates around late-type stars \cite[e.g., a few percent; see Figure 8 in][]{2013ApJ...771...18G}. 

%We checked whether the material causing the dips could account for the observed infrared excess, but found we could not rule out a spherical dust cloud where much of the material never passes in front of the star.

%======================= CONCLUSION =========================

\section{SUMMARY}
\label{sec-conclusion}

We presented disk inclinations for the three known dippers with resolved sub-mm data. We found disks with a range of viewing geometries, most notably the face-on transition disk J1604-2130. Our findings show that nearly edge-on disk inclinations are {\em not} a defining characteristic of the dippers, contrary to the currently proposed mechanisms explaining the dips, suggesting that additional models should be explored. Resolving more dipper disks with ALMA, and exploring techniques such as spectroastrometry that can directly probe the inner disk \cite[e.g.][]{2008ApJ...684.1323P}, will be essential to understanding and properly interpreting the dippers in the context of planet formation.

%======================= ACKNOWLEDGEMENTS =========================

\section*{Acknowledgements}
Support comes from NSF grant AST-1208911 (MA), NASA grants NNX11AC33G (EG) \& NNX15AC92G (JPW), ERC grant 279973 (MCW), Hubble Fellowship 51364 (AWM), and a Royal Society University Research Fellowship (GK).

%%%%%%%%%%%%%%%%%%%%%%%%%%%%%%%%%%%%%%%%%%%%%%%%%%
%%%%%%%%%%%%%%%%%%%% REFERENCES %%%%%%%%%%%%%%%%%%
%%%%%%%%%%%%%%%%%%%%%%%%%%%%%%%%%%%%%%%%%%%%%%%%%%

\bibliographystyle{mnras}
%\bibliography{../../bib.bib}

\begin{thebibliography}{}
\makeatletter
\relax
\def\mn@urlcharsother{\let\do\@makeother \do\$\do\&\do\#\do\^\do\_\do\%\do\~}
\def\mn@doi{\begingroup\mn@urlcharsother \@ifnextchar [ {\mn@doi@}
  {\mn@doi@[]}}
\def\mn@doi@[#1]#2{\def\@tempa{#1}\ifx\@tempa\@empty \href
  {http://dx.doi.org/#2} {doi:#2}\else \href {http://dx.doi.org/#2} {#1}\fi
  \endgroup}
\def\mn@eprint#1#2{\mn@eprint@#1:#2::\@nil}
\def\mn@eprint@arXiv#1{\href {http://arxiv.org/abs/#1} {{\tt arXiv:#1}}}
\def\mn@eprint@dblp#1{\href {http://dblp.uni-trier.de/rec/bibtex/#1.xml}
  {dblp:#1}}
\def\mn@eprint@#1:#2:#3:#4\@nil{\def\@tempa {#1}\def\@tempb {#2}\def\@tempc
  {#3}\ifx \@tempc \@empty \let \@tempc \@tempb \let \@tempb \@tempa \fi \ifx
  \@tempb \@empty \def\@tempb {arXiv}\fi \@ifundefined
  {mn@eprint@\@tempb}{\@tempb:\@tempc}{\expandafter \expandafter \csname
  mn@eprint@\@tempb\endcsname \expandafter{\@tempc}}}

\bibitem[\protect\citeauthoryear{{Alencar} et~al.,}{{Alencar}
  et~al.}{2010}]{2010A&A...519A..88A}
{Alencar} S.~H.~P.,  et~al., 2010, \mn@doi [\aap]
  {10.1051/0004-6361/201014184}, \href
  {http://adsabs.harvard.edu/abs/2010A%26A...519A..88A} {519, A88}

\bibitem[\protect\citeauthoryear{{Andrews} \& {Williams}}{{Andrews} \&
  {Williams}}{2007}]{2007ApJ...671.1800A}
{Andrews} S.~M.,  {Williams} J.~P.,  2007, \mn@doi [\apj] {10.1086/522885},
  \href {http://adsabs.harvard.edu/abs/2007ApJ...671.1800A} {671, 1800}

\bibitem[\protect\citeauthoryear{{Ansdell} et~al.,}{{Ansdell}
  et~al.}{2016}]{2016ApJ...816...69A}
{Ansdell} M.,  et~al., 2016, \mn@doi [\apj] {10.3847/0004-637X/816/2/69}, \href
  {http://adsabs.harvard.edu/abs/2016ApJ...816...69A} {816, 69}

\bibitem[\protect\citeauthoryear{{Biller} et~al.,}{{Biller}
  et~al.}{2012}]{2012ApJ...753L..38B}
{Biller} B.,  et~al., 2012, \mn@doi [\apjl] {10.1088/2041-8205/753/2/L38},
  \href {http://adsabs.harvard.edu/abs/2012ApJ...753L..38B} {753, L38}

\bibitem[\protect\citeauthoryear{{Bodman} et~al.,}{{Bodman}
  et~al.}{2016}]{2016arXiv160503985B}
{Bodman} E.~H.~L.,  et~al., 2016, preprint, \href
  {http://adsabs.harvard.edu/abs/2016arXiv160503985B} {} (\mn@eprint {arXiv}
  {1605.03985})

\bibitem[\protect\citeauthoryear{{Bouvier} et~al.,}{{Bouvier}
  et~al.}{1999}]{1999A&A...349..619B}
{Bouvier} J.,  et~al., 1999, \aap, \href
  {http://adsabs.harvard.edu/abs/1999A%26A...349..619B} {349, 619}

\bibitem[\protect\citeauthoryear{{Carpenter}, {Ricci}  \& {Isella}}{{Carpenter}
  et~al.}{2014}]{2014ApJ...787...42C}
{Carpenter} J.~M.,  {Ricci} L.,   {Isella} A.,  2014, \mn@doi [\apj]
  {10.1088/0004-637X/787/1/42}, \href
  {http://adsabs.harvard.edu/abs/2014ApJ...787...42C} {787, 42}

\bibitem[\protect\citeauthoryear{{Casassus}, {Perez M.}, {Jord{\'a}n},
  {M{\'e}nard}, {Cuadra}, {Schreiber}, {Hales}  \& {Ercolano}}{{Casassus}
  et~al.}{2012}]{2012ApJ...754L..31C}
{Casassus} S.,  {Perez M.} S.,  {Jord{\'a}n} A.,  {M{\'e}nard} F.,  {Cuadra}
  J.,  {Schreiber} M.~R.,  {Hales} A.~S.,   {Ercolano} B.,  2012, \mn@doi
  [\apjl] {10.1088/2041-8205/754/2/L31}, \href
  {http://adsabs.harvard.edu/abs/2012ApJ...754L..31C} {754, L31}

\bibitem[\protect\citeauthoryear{{Casassus} et~al.,}{{Casassus}
  et~al.}{2013}]{2013Natur.493..191C}
{Casassus} S.,  et~al., 2013, \mn@doi [\nat] {10.1038/nature11769}, \href
  {http://adsabs.harvard.edu/abs/2013Natur.493..191C} {493, 191}

\bibitem[\protect\citeauthoryear{{Close} et~al.,}{{Close}
  et~al.}{2014}]{2014ApJ...781L..30C}
{Close} L.~M.,  et~al., 2014, \mn@doi [\apjl] {10.1088/2041-8205/781/2/L30},
  \href {http://adsabs.harvard.edu/abs/2014ApJ...781L..30C} {781, L30}

\bibitem[\protect\citeauthoryear{{Cody} et~al.,}{{Cody}
  et~al.}{2014}]{2014AJ....147...82C}
{Cody} A.~M.,  et~al., 2014, \mn@doi [\aj] {10.1088/0004-6256/147/4/82}, \href
  {http://adsabs.harvard.edu/abs/2014AJ....147...82C} {147, 82}

\bibitem[\protect\citeauthoryear{{Dahm}}{{Dahm}}{2010}]{2010AJ....140.1444D}
{Dahm} S.~E.,  2010, \mn@doi [\aj] {10.1088/0004-6256/140/5/1444}, \href
  {http://adsabs.harvard.edu/abs/2010AJ....140.1444D} {140, 1444}

\bibitem[\protect\citeauthoryear{{Dahm} \& {Carpenter}}{{Dahm} \&
  {Carpenter}}{2009}]{2009AJ....137.4024D}
{Dahm} S.~E.,  {Carpenter} J.~M.,  2009, \mn@doi [\aj]
  {10.1088/0004-6256/137/4/4024}, \href
  {http://adsabs.harvard.edu/abs/2009AJ....137.4024D} {137, 4024}

\bibitem[\protect\citeauthoryear{{Dahm}, {Slesnick}  \& {White}}{{Dahm}
  et~al.}{2012}]{2012ApJ...745...56D}
{Dahm} S.~E.,  {Slesnick} C.~L.,   {White} R.~J.,  2012, \mn@doi [\apj]
  {10.1088/0004-637X/745/1/56}, \href
  {http://cdsads.u-strasbg.fr/abs/2012ApJ...745...56D} {745, 56}

\bibitem[\protect\citeauthoryear{{Facchini}, {Ricci}  \& {Lodato}}{{Facchini}
  et~al.}{2014}]{2014MNRAS.442.3700F}
{Facchini} S.,  {Ricci} L.,   {Lodato} G.,  2014, \mn@doi [\mnras]
  {10.1093/mnras/stu1149}, \href
  {http://adsabs.harvard.edu/abs/2014MNRAS.442.3700F} {442, 3700}

\bibitem[\protect\citeauthoryear{{Fukagawa}, {Tamura}, {Itoh}, {Kudo},
  {Imaeda}, {Oasa}, {Hayashi}  \& {Hayashi}}{{Fukagawa}
  et~al.}{2006}]{2006ApJ...636L.153F}
{Fukagawa} M.,  {Tamura} M.,  {Itoh} Y.,  {Kudo} T.,  {Imaeda} Y.,  {Oasa} Y.,
  {Hayashi} S.~S.,   {Hayashi} M.,  2006, \mn@doi [\apjl] {10.1086/500128},
  \href {http://adsabs.harvard.edu/abs/2006ApJ...636L.153F} {636, L153}

\bibitem[\protect\citeauthoryear{{Gaidos}, {Fischer}, {Mann}  \&
  {Howard}}{{Gaidos} et~al.}{2013}]{2013ApJ...771...18G}
{Gaidos} E.,  {Fischer} D.~A.,  {Mann} A.~W.,   {Howard} A.~W.,  2013, \mn@doi
  [\apj] {10.1088/0004-637X/771/1/18}, \href
  {http://adsabs.harvard.edu/abs/2013ApJ...771...18G} {771, 18}

\bibitem[\protect\citeauthoryear{{Ho}, {Moran}  \& {Lo}}{{Ho}
  et~al.}{2004}]{2004ApJ...616L...1H}
{Ho} P.~T.~P.,  {Moran} J.~M.,   {Lo} K.~Y.,  2004, \mn@doi [\apjl]
  {10.1086/423245}, \href {http://adsabs.harvard.edu/abs/2004ApJ...616L...1H}
  {616, L1}

\bibitem[\protect\citeauthoryear{{Howard} et~al.,}{{Howard}
  et~al.}{2010}]{2010Sci...330..653H}
{Howard} A.~W.,  et~al., 2010, \mn@doi [Science] {10.1126/science.1194854},
  \href {http://adsabs.harvard.edu/abs/2010Sci...330..653H} {330, 653}

\bibitem[\protect\citeauthoryear{{Howell} et~al.,}{{Howell}
  et~al.}{2014}]{2014PASP..126..398H}
{Howell} S.~B.,  et~al., 2014, \mn@doi [\pasp] {10.1086/676406}, \href
  {http://adsabs.harvard.edu/abs/2014PASP..126..398H} {126, 398}

\bibitem[\protect\citeauthoryear{{Ireland}, {Kraus}, {Martinache}, {Law}  \&
  {Hillenbrand}}{{Ireland} et~al.}{2011}]{2011ApJ...726..113I}
{Ireland} M.~J.,  {Kraus} A.,  {Martinache} F.,  {Law} N.,   {Hillenbrand}
  L.~A.,  2011, \mn@doi [\apj] {10.1088/0004-637X/726/2/113}, \href
  {http://adsabs.harvard.edu/abs/2011ApJ...726..113I} {726, 113}

\bibitem[\protect\citeauthoryear{{Kraus}, {Ireland}, {Martinache}  \&
  {Lloyd}}{{Kraus} et~al.}{2008}]{2008ApJ...679..762K}
{Kraus} A.~L.,  {Ireland} M.~J.,  {Martinache} F.,   {Lloyd} J.~P.,  2008,
  \mn@doi [\apj] {10.1086/587435}, \href
  {http://adsabs.harvard.edu/abs/2008ApJ...679..762K} {679, 762}

\bibitem[\protect\citeauthoryear{{Krijt} \& {Dominik}}{{Krijt} \&
  {Dominik}}{2011}]{2011A&A...531A..80K}
{Krijt} S.,  {Dominik} C.,  2011, \mn@doi [\aap] {10.1051/0004-6361/201116757},
  \href {http://adsabs.harvard.edu/abs/2011A%26A...531A..80K} {531, A80}

\bibitem[\protect\citeauthoryear{{Manara}, {Testi}, {Natta}  \&
  {Alcal{\'a}}}{{Manara} et~al.}{2015}]{2015A&A...579A..66M}
{Manara} C.~F.,  {Testi} L.,  {Natta} A.,   {Alcal{\'a}} J.~M.,  2015, \mn@doi
  [\aap] {10.1051/0004-6361/201526169}, \href
  {http://cdsads.u-strasbg.fr/abs/2015A%26A...579A..66M} {579, A66}

\bibitem[\protect\citeauthoryear{{Marino}, {Perez}  \& {Casassus}}{{Marino}
  et~al.}{2015}]{2015ApJ...798L..44M}
{Marino} S.,  {Perez} S.,   {Casassus} S.,  2015, \mn@doi [\apjl]
  {10.1088/2041-8205/798/2/L44}, \href
  {http://adsabs.harvard.edu/abs/2015ApJ...798L..44M} {798, L44}

\bibitem[\protect\citeauthoryear{{Mathews}, {Williams}  \&
  {M{\'e}nard}}{{Mathews} et~al.}{2012}]{2012ApJ...753...59M}
{Mathews} G.~S.,  {Williams} J.~P.,   {M{\'e}nard} F.,  2012, \mn@doi [\apj]
  {10.1088/0004-637X/753/1/59}, \href
  {http://adsabs.harvard.edu/abs/2012ApJ...753...59M} {753, 59}

\bibitem[\protect\citeauthoryear{{McClure} et~al.,}{{McClure}
  et~al.}{2010}]{2010ApJS..188...75M}
{McClure} M.~K.,  et~al., 2010, \mn@doi [\apjs] {10.1088/0067-0049/188/1/75},
  \href {http://cdsads.u-strasbg.fr/abs/2010ApJS..188...75M} {188, 75}

\bibitem[\protect\citeauthoryear{{McGinnis} et~al.,}{{McGinnis}
  et~al.}{2015}]{2015A&A...577A..11M}
{McGinnis} P.~T.,  et~al., 2015, \mn@doi [\aap] {10.1051/0004-6361/201425475},
  \href {http://adsabs.harvard.edu/abs/2015A%26A...577A..11M} {577, A11}

\bibitem[\protect\citeauthoryear{{McMullin}, {Waters}, {Schiebel}, {Young}  \&
  {Golap}}{{McMullin} et~al.}{2007}]{2007ASPC..376..127M}
{McMullin} J.~P.,  {Waters} B.,  {Schiebel} D.,  {Young} W.,   {Golap} K.,
  2007, in {Shaw} R.~A.,  {Hill} F.,   {Bell} D.~J.,  eds,  Astronomical
  Society of the Pacific Conference Series Vol. 376, Astronomical Data Analysis
  Software and Systems XVI. p.~127

\bibitem[\protect\citeauthoryear{{Morales-Calder{\'o}n}
  et~al.,}{{Morales-Calder{\'o}n} et~al.}{2011}]{2011ApJ...733...50M}
{Morales-Calder{\'o}n} M.,  et~al., 2011, \mn@doi [\apj]
  {10.1088/0004-637X/733/1/50}, \href
  {http://adsabs.harvard.edu/abs/2011ApJ...733...50M} {733, 50}

\bibitem[\protect\citeauthoryear{{Natta}, {Testi}, {Comer{\'o}n}, {Oliva},
  {D'Antona}, {Baffa}, {Comoretto}  \& {Gennari}}{{Natta}
  et~al.}{2002}]{2002A&A...393..597N}
{Natta} A.,  {Testi} L.,  {Comer{\'o}n} F.,  {Oliva} E.,  {D'Antona} F.,
  {Baffa} C.,  {Comoretto} G.,   {Gennari} S.,  2002, \mn@doi [\aap]
  {10.1051/0004-6361:20021065}, \href
  {http://adsabs.harvard.edu/abs/2002A%26A...393..597N} {393, 597}

\bibitem[\protect\citeauthoryear{{Natta}, {Testi}, {Muzerolle}, {Randich},
  {Comer{\'o}n}  \& {Persi}}{{Natta} et~al.}{2004}]{2004A&A...424..603N}
{Natta} A.,  {Testi} L.,  {Muzerolle} J.,  {Randich} S.,  {Comer{\'o}n} F.,
  {Persi} P.,  2004, \mn@doi [\aap] {10.1051/0004-6361:20040356}, \href
  {http://adsabs.harvard.edu/abs/2004A%26A...424..603N} {424, 603}

\bibitem[\protect\citeauthoryear{{Pecaut}, {Mamajek}  \& {Bubar}}{{Pecaut}
  et~al.}{2012}]{2012ApJ...746..154P}
{Pecaut} M.~J.,  {Mamajek} E.~E.,   {Bubar} E.~J.,  2012, \mn@doi [\apj]
  {10.1088/0004-637X/746/2/154}, \href
  {http://adsabs.harvard.edu/abs/2012ApJ...746..154P} {746, 154}

\bibitem[\protect\citeauthoryear{{Petigura}, {Marcy}  \& {Howard}}{{Petigura}
  et~al.}{2013}]{2013ApJ...770...69P}
{Petigura} E.~A.,  {Marcy} G.~W.,   {Howard} A.~W.,  2013, \mn@doi [\apj]
  {10.1088/0004-637X/770/1/69}, \href
  {http://adsabs.harvard.edu/abs/2013ApJ...770...69P} {770, 69}

\bibitem[\protect\citeauthoryear{{Phan-Bao} et~al.,}{{Phan-Bao}
  et~al.}{2008}]{2008ApJ...689L.141P}
{Phan-Bao} N.,  et~al., 2008, \mn@doi [\apjl] {10.1086/595961}, \href
  {http://adsabs.harvard.edu/abs/2008ApJ...689L.141P} {689, L141}

\bibitem[\protect\citeauthoryear{{Pontoppidan}, {Blake}, {van Dishoeck},
  {Smette}, {Ireland}  \& {Brown}}{{Pontoppidan}
  et~al.}{2008}]{2008ApJ...684.1323P}
{Pontoppidan} K.~M.,  {Blake} G.~A.,  {van Dishoeck} E.~F.,  {Smette} A.,
  {Ireland} M.~J.,   {Brown} J.,  2008, \mn@doi [\apj] {10.1086/590400}, \href
  {http://adsabs.harvard.edu/abs/2008ApJ...684.1323P} {684, 1323}

\bibitem[\protect\citeauthoryear{{Ricci}, {Testi}, {Natta}, {Scholz}  \& {de
  Gregorio-Monsalvo}}{{Ricci} et~al.}{2012}]{2012ApJ...761L..20R}
{Ricci} L.,  {Testi} L.,  {Natta} A.,  {Scholz} A.,   {de Gregorio-Monsalvo}
  I.,  2012, \mn@doi [\apjl] {10.1088/2041-8205/761/2/L20}, \href
  {http://adsabs.harvard.edu/abs/2012ApJ...761L..20R} {761, L20}

\bibitem[\protect\citeauthoryear{{Rodigas}, {Follette}, {Weinberger}, {Close}
  \& {Hines}}{{Rodigas} et~al.}{2014}]{2014ApJ...791L..37R}
{Rodigas} T.~J.,  {Follette} K.~B.,  {Weinberger} A.,  {Close} L.,   {Hines}
  D.~C.,  2014, \mn@doi [\apjl] {10.1088/2041-8205/791/2/L37}, \href
  {http://adsabs.harvard.edu/abs/2014ApJ...791L..37R} {791, L37}

\bibitem[\protect\citeauthoryear{{Silburt}, {Gaidos}  \& {Wu}}{{Silburt}
  et~al.}{2015}]{2015ApJ...799..180S}
{Silburt} A.,  {Gaidos} E.,   {Wu} Y.,  2015, \mn@doi [\apj]
  {10.1088/0004-637X/799/2/180}, \href
  {http://adsabs.harvard.edu/abs/2015ApJ...799..180S} {799, 180}

\bibitem[\protect\citeauthoryear{{Skrutskie} et~al.,}{{Skrutskie}
  et~al.}{2006}]{2006AJ....131.1163S}
{Skrutskie} M.~F.,  et~al., 2006, \mn@doi [\aj] {10.1086/498708}, \href
  {http://adsabs.harvard.edu/abs/2006AJ....131.1163S} {131, 1163}

\bibitem[\protect\citeauthoryear{{Takami} et~al.,}{{Takami}
  et~al.}{2014}]{2014ApJ...795...71T}
{Takami} M.,  et~al., 2014, \mn@doi [\apj] {10.1088/0004-637X/795/1/71}, \href
  {http://cdsads.u-strasbg.fr/abs/2014ApJ...795...71T} {795, 71}

\bibitem[\protect\citeauthoryear{{Vanderburg} \& {Johnson}}{{Vanderburg} \&
  {Johnson}}{2014}]{2014PASP..126..948V}
{Vanderburg} A.,  {Johnson} J.~A.,  2014, \mn@doi [\pasp] {10.1086/678764},
  \href {http://adsabs.harvard.edu/abs/2014PASP..126..948V} {126, 948}

\bibitem[\protect\citeauthoryear{{Verhoeff} et~al.,}{{Verhoeff}
  et~al.}{2011}]{2011A&A...528A..91V}
{Verhoeff} A.~P.,  et~al., 2011, \mn@doi [\aap] {10.1051/0004-6361/201014952},
  \href {http://adsabs.harvard.edu/abs/2011A%26A...528A..91V} {528, A91}

\bibitem[\protect\citeauthoryear{{Whelan}, {Ray}, {Bacciotti}, {Natta}, {Testi}
   \& {Randich}}{{Whelan} et~al.}{2005}]{2005Natur.435..652W}
{Whelan} E.~T.,  {Ray} T.~P.,  {Bacciotti} F.,  {Natta} A.,  {Testi} L.,
  {Randich} S.,  2005, \mn@doi [\nat] {10.1038/nature03598}, \href
  {http://adsabs.harvard.edu/abs/2005Natur.435..652W} {435, 652}

\bibitem[\protect\citeauthoryear{{Zhang}, {Isella}, {Carpenter}  \&
  {Blake}}{{Zhang} et~al.}{2014}]{2014ApJ...791...42Z}
{Zhang} K.,  {Isella} A.,  {Carpenter} J.~M.,   {Blake} G.~A.,  2014, \mn@doi
  [\apj] {10.1088/0004-637X/791/1/42}, \href
  {http://adsabs.harvard.edu/abs/2014ApJ...791...42Z} {791, 42}

\bibitem[\protect\citeauthoryear{{van der Marel}, {van Dishoeck}, {Bruderer},
  {P{\'e}rez}  \& {Isella}}{{van der Marel} et~al.}{2015}]{2015A&A...579A.106V}
{van der Marel} N.,  {van Dishoeck} E.~F.,  {Bruderer} S.,  {P{\'e}rez} L.,
  {Isella} A.,  2015, \mn@doi [\aap] {10.1051/0004-6361/201525658}, \href
  {http://adsabs.harvard.edu/abs/2015A%26A...579A.106V} {579, A106}

\makeatother
\end{thebibliography}

% Don't change these lines
\bsp	% typesetting comment
\label{lastpage}
\end{document}